\documentclass{aa}

\usepackage{psfig}
\usepackage{graphics}

\def\Ni{\noindent}
\def\msun{$M_{\odot}$}

\def\etal{{\it et al.}}
\def\ros{{\it ROSAT}}
\def\grad{$^\circ$}
\def\fss{\hbox{$.\!\!^{\rm s}$}}        
\def\asec{\ifmmode ^{\prime\prime}\else$^{\prime\prime}$\fi}
\def\amin{\ifmmode ^{\prime}\else$^{\prime}$\fi}
\def\D{\ifmmode ^{\circ}\else$^{\circ}$\fi}
\def\H{$^{\rm h}$}\def\M{$^{\rm m}$}
\newbox\grsign \setbox\grsign=\hbox{$>$}
\newdimen\grdimen \grdimen=\ht\grsign
\newbox\laxbox \newbox\gaxbox
\setbox\gaxbox=\hbox{\raise.5ex\hbox{$>$}\llap
     {\lower.5ex\hbox{$\sim$}}}\ht1=\grdimen\dp1=0pt
\setbox\laxbox=\hbox{\raise.5ex\hbox{$<$}\llap
     {\lower.5ex\hbox{$\sim$}}}\ht2=\grdimen\dp2=0pt
\def\gax{\mathrel{\copy\gaxbox}}
\def\lax{\mathrel{\copy\laxbox}}
\def\h{$^{\rm h}$}\def\m{$^{\rm m}$}

\def\mdot{$\dot M$}
\def\msun{$M_{\odot}$}

\def\v7{V751\,Cyg}

\def\0927{RX\,J0927.7--4758}
\def\rxj0513{RX\,J0513.9--6951}
\font\pr=cmr7

\def\II{{\pr II}}

\begin{document}

   \thesaurus{06         
              (02.01.2;  
               08.09.2;  
               08.14.2;  
               13.25.5)} 

\title{Transient supersoft X-ray emission from V751 Cyg during the
  optical low-state}

\author{J. Greiner\inst{1}, G.H. Tovmassian\inst{2},  
  R. Di\,Stefano\inst{3}, A. Prestwich\inst{3}, R. Gonz\'alez-Riestra\inst{4},
  L. Szentasko\inst{5}, C. Chavarr\'{\i}a\inst{2}}

\offprints{J.\,Greiner,\,jgreiner@aip.de}

  \institute{$^1$ Astrophysical Institute Potsdam, An der Sternwarte 16,
             14482 Potsdam, Germany \\
        $^2$ Observatorio Astronomico Nacional, Instituto de Astronom\'{\i}a, 
       UNAM, AP 877, 22860, Ensenada, B.C., M\'{e}xico \\
        $^3$ Harvard-Smithonian Center for Astrophysics, 60 Garden Street, 
               Cambridge, MA 02138, USA \\
         $^4$ Laboratorio de Astrof\'{\i}sica Espacial y F\'{\i}sica 
              Fundamental, P.O. Box 50727,  28080 Madrid, Spain \\
        $^5$ Budapest, Hungary \\
       }

   \date{Received 27 July 1998; accepted 2 December 1998}
 
   \authorrunning{Greiner et al.}
   \titlerunning{Transient supersoft X-ray emission from \v7}
   \maketitle

   \markboth{Greiner et al.}{Transient supersoft X-ray emission from V751 Cyg}

\begin{abstract}
We have observed \v7\ with the ROSAT HRI in a target-of-opportunity mode 
during its recent optical low state and clearly detect it at a count rate of 
0.015 cts/s. The X-ray intensity is a factor of 7--19 (depending on the exact 
X-ray spectral shape) higher than the upper limit obtained with the
ROSAT PSPC during the optical high state, thus suggesting an anti-correlation
of X-ray and optical intensity. Spectral investigations suggest a very soft
X-ray spectrum. We investigate archival IUE data of \v7\ and derive
a distance of \v7\ of d$\approx$500 pc based on the extinction
estimate of $E(B-V)$=0.25$\pm$0.05. This implies that the X-ray emission 
is very luminous, on the order of 10$^{34}$--10$^{36}$ erg/s.

We have obtained quasi-simultaneous optical photometry and spectroscopy.
The spectrum during the optical low-state is characterized by a very
blue continuum and the presence of strong emission lines of the Balmer 
series and HeI. Also, HeII 4686 \AA\ is clearly detected.

We establish that \v7\ is
a transient supersoft X-ray source and speculate that other
VY Scl stars may also be of similar type.

\end{abstract}


\section{Introduction}

Supersoft X-ray binaries (SSBs)  
have been established as a new class by ROSAT observations over
the past several years (e.g. Greiner 1996 and references therein).
Supersoft X-ray binaries are characterized by luminous 
($10^{36}-10^{38}$ erg s$^{-1}$) emission with typical temperatures 
of 20--40 eV, and are thought to contain white dwarfs accreting mass at
rates sufficiently high to allow 
quasi-steady nuclear burning of the accreted
matter (van den Heuvel \etal\ 1992). 
Although the estimated Galactic population is 
large (Di\,Stefano \& Rappaport 1994), 
only a few sources in the Milky Way are known, because of the effect
of interstellar absorption. Indeed, 
most of the known supersoft X-ray binaries (SSB) are located in external
galaxies  making detailed optical observations difficult.

The long-term monitoring of the transient supersoft X-ray source
RX J0513.9-6951 (Schaeidt \etal\ 1993) has revealed quasi-periodic 
optical intensity dips of about 4 weeks duration (Reinsch \etal\ 1996,
Southwell \etal\ 1996) which occur simultaneous to the X-ray on states. 
The similarity of the optical behaviour of this supersoft transient with
that of VY Scl stars led us to  speculate that some of the VY Scl stars may 
indeed be hitherto unrecognized supersoft transients. If true, this would allow
us to search for and find new supersoft X-ray sources independent of X-ray 
surveys. 
 This is important since the estimated population in the Galaxy 
is large (Di\,Stefano \& Rappaport 1994), but only few systems are known.
We have therefore systematically checked the 14 known 
VY Scl stars for their optical state over the last two years in order to carry
out soft X-ray observations once a VY Scl star  enters an optical low state.
In this paper we report on the results of some of these observations,
which indicate that at least one VY Scl star,
\v7, is a supersoft X-ray binary. 

We begin below with a brief introduction to the properties of VY Scl stars in 
general, and \v7\ in particular. We then go on in \S 2  and \S 3 to describe, 
respectively, our observations (Tabs. \ref{xlog}, \ref{olog}) and results. 
In \S 4 we discuss \v7, and address in \S 5 the question of whether other 
VY Scl stars may also be SSBs. In \S 6 we summarize our conclusions
(see also Greiner \& Di\,Stefano 1998).

\begin{table*}
\caption{Log of ROSAT observations of \v7}
\vspace{-0.25cm}
\begin{tabular}{ccccccc}
\hline
\noalign{\smallskip}
  Date  & JD & Detector & Exposure & Offaxis Angle &  CR$^1$ & Opt. State \\
        &    &          & (sec)  & (\amin)       & (cts/s) & \\
\noalign{\smallskip}
\hline
\noalign{\smallskip}
  Nov. 19/20, 1990 & 2448216 & PSPC & ~\,~\,370 & 0--55 & $<$ 0.019  &  high \\
  Nov. 3, 1992     & 2448930 & PSPC & ~\,3637 & 52    & $<$ 0.0058 &   high \\
  Jun. 3, 1997     & 2450603 & HRI & ~\,4663 & 0.3   & 0.015$\pm$0.002 & low \\
  Dec. 2--8, 1997  & 2450788 & HRI &   10814 & 0.2   & 0.010$\pm$0.001 & low \\
\noalign{\smallskip}
\hline
\end{tabular}

\noindent{\Ni\small 
 $^{(1)}$ The values for the two PSPC observations are 3$\sigma$ upper limits.
           Note that the 0.015 HRI cts/s correspond to a PSPC rate of 
           0.039--0.11 cts/s (see section 3.5).}
\label{xlog}
\end{table*}

\begin{table*}[ht]
\caption{Log of Optical Observations}
\vspace{-0.25cm}
\begin{tabular}{ccccrrl}
      \hline
      \noalign{\smallskip}
Date & JD & Telescope + Equip. & Filter/Wavelength & Duration & Exp. & Site \\
     &    &                    &                &  (min)     & (sec) &   \\ 
 \noalign{\smallskip}
 \hline
 \noalign{\smallskip}
Sep. 28, 1997& 2450719 & 2.1m, B\&Ch sp. & 3600--7500 \AA & 60~~~ & 600 & SPM\\
Sep. 29, 1997& 2450720 & 2.1m, B\&Ch sp. & 4290--8385 \AA & 30~~~ & 600 & SPM\\
Sep. 30, 1997& 2450721 & 2.1m, B\&Ch sp. & 5460--8530 \AA & 80~~~ & 1200 & SPM\\
Sep. 30, 1997& 2450721 & 1.5m, Danish Phot.& {\sl uvby}-$\beta$& 5~~~ &10 &SPM\\
Dec. 03, 1997 & 2450785 & 1.5m, CCD direct& UBVRI & 80~~~ & 180-300 & SPM\\
       \noalign{\smallskip}
      \hline
   \end{tabular}
   \label{olog}
   \end{table*}

VY Scl stars (or sometimes called anti-dwarf novae)
are bright most of the time, but occasionally drop in brightness 
at irregular intervals by 2--8 magnitudes. The transitions between the 
brightness levels take place on a time scale of days to weeks. 
During the optical low states the optical Balmer emission lines become very
narrow and show recombination ratios, the UV emission lines disappear and 
occasionally the infrared spectrum reveals the signatures of the cool donor
star (see e.g. Warner 1995). These observations suggest that most or even
all of the accretion disk disappears during the low state.
VY Scl  variables are typically found at $P >$ 3 hrs (i.e. exclusively 
above the period gap), and with large mass transfer rate 
\mdot $\sim$ 10$^{-8}$ \msun/yr in the optical high state (upper right corner 
of the $P_{\rm orb}-$\mdot\ diagram of
Osaki 1995), and thus are thought to be steady accretors (or dwarf novae in a 
state of continuous eruption as suggested by Kraft 1964) with hot disks.
Their disks are thus assumed to be thermally and tidally stable.

\v7\  (= EM* LkHA 170 = SVS 1202) was discovered by Martynov (1958) and
originally classified as an R Coronae Borealis
star due to its irregular variations between 13.5--14.0 mag (pg) and
the occasional fading down to about 16 mag (pg) (Martynov \& Kholopov 1958).
Further observations revealed additional strong fading events (Wenzel 1963),
but spectroscopic results, most notably the lack of absorption lines and
the continuous and very blue spectrum (Herbig 1958,
Herbig \& Rao 1972, Downes \etal\ 1995), are clearly inconsistent with the
interpretation of a carbon-rich, pulsating late supergiant.
Now-a-days \v7\ is classified as a nova-like variable based
on the observed rapid flickering and the complete similarity to VY Scl
(Robinson \etal\ 1974).

\v7\ has poorly determined binary parameter properties, though its 
classification as a VY Scl star is beyond doubt (Wenzel 1963, Robinson \etal\ 
1974, Downes \etal\ 1995).
Both the small reddening implied by the blue colors 
($U-B\approx -1.02, B-V\approx -0.12$; see Burrel \& Mould 1973, 
Munari \etal\ 1997) and the projected vicinity to both IC 5070 (920 pc)
and the $^{13}$CO cloud \#34 (Dobashi \etal\ 1994; 800 pc) suggest that \v7\
is at 800 pc distance at maximum. With these assumptions, i.e. no extinction
and $d < 920$pc it has been deduced 
that \v7\ is fainter than $M_{\rm V}\approx +4.8$ at maximum
light, and fainter than $M_{\rm V}\approx +7.0$ at minimum light 
(Robinson \etal\ 1974). The orbital period of \v7\ is not well known, the only 
report being that of Bell \& Walker (1980) on $P \approx$ 0\fd25.

Livio \& Pringle (1994) proposed a model for the group of VY Scl stars in which
optical low states are associated with a reduced mass transfer rate caused 
by a magnetic spot covering 
temporarily the $L_1$ region. This mechanism works predominantly at short 
orbital periods because the level of magnetic activity increases with the 
rotation rate of the star (which in turn is coupled to the orbit).

The above described model to explain VY Scl star low states
may also work for supersoft X-ray sources (Alcock \etal\
1997) with their longer periods because the magnetic activity actually scales 
like $P_{\rm rot}/\tau_{rm c}$ (= Rossby number) with $\tau_{rm c}$ 
being the convective overturn time in the envelope (Schrijver 1994). 
And $\tau_{rm c}$ is longer for stars with a deeper convective envelope, 
precisely what is expected for the evolved secondaries in supersoft X-ray 
sources according to the standard model (van den Heuvel \etal\ 1992, 
DiStefano \& Rappaport 1994).

\section{Observations}

   \begin{figure*}
    \vspace{-0.35cm}
    \vbox{\psfig{figure=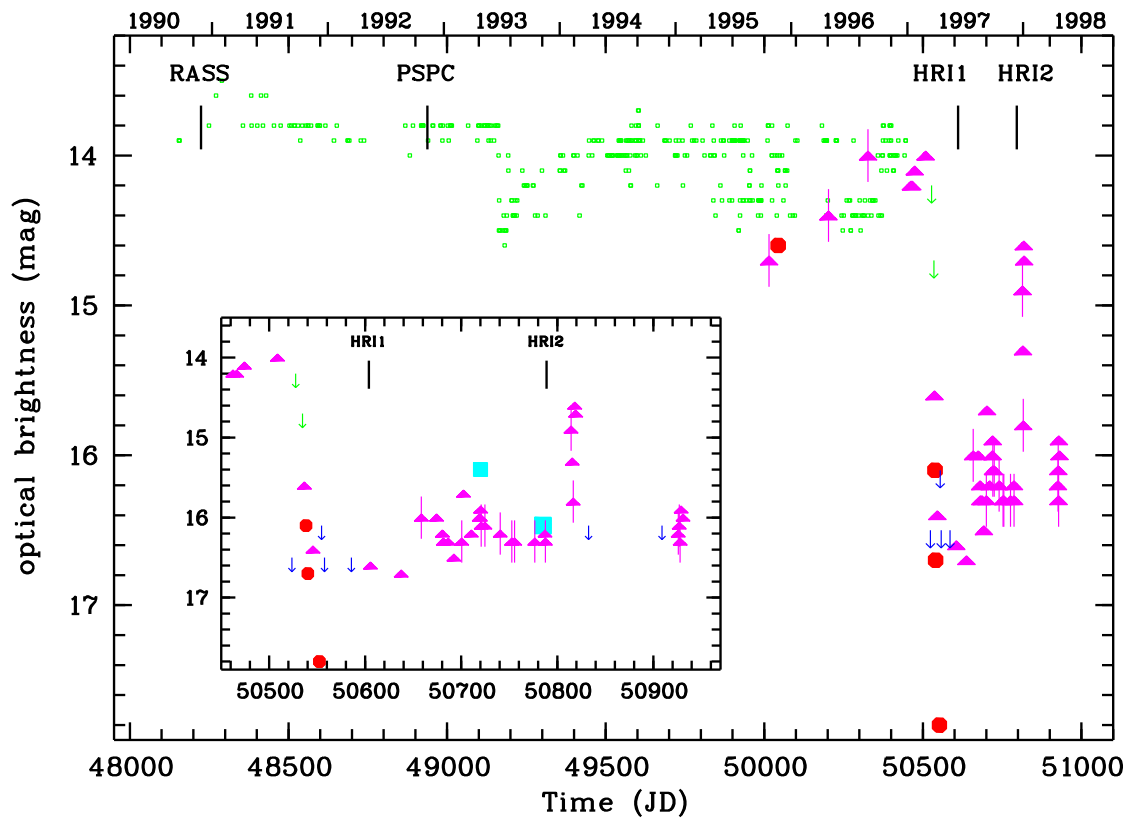,width=14.9cm,%
          bbllx=2.cm,bblly=1.9cm,bburx=17.7cm,bbury=13.5cm,clip=}}\par
    \vspace{-0.15cm}
    \caption[pha]{Optical light curve of V751 Cyg over the last 8 years:
         small (gray) squares denote measurements as reported to AFOEV 
        and triangles denote measurements as reported to VSNET (uncertain
        measurements are plotted with an error bar while the error of the
        certain measurements is about the size of the symbol sign). 
       Large filled circles
        are CCD measurements of the Ouda team (also taken from VSNET).
        Arrows  denote  upper limits.
         At the top the times of the  ROSAT observations are marked.
         V751 Cyg is detected only in the HRI observations during the optical
        low state. The inset shows a blow-up of the optical low-state together
       with the times of the two HRI observations. The two squares in the inset
        represent the mean brightness on Sep. 29/30, 1997 as derived from
        our spectra and our Dec. 3, 1997 photometry (at the time of the second 
       HRI observation).
        }
      \label{lc}
   \end{figure*}

\subsection{X-ray observations}

After notification of the fading of \v7\ in March 1997 we performed a
target-of-opportunity HRI observation (after the opening of the ROSAT 
observing window) for a total of 4660 sec on June 3, 1997.
We detect 12 X-ray sources above 4$\sigma$ in the whole HRI field of view
(Tab. \ref{xlist}) one of which coincides within 1\asec\ with \v7.
This X-ray source, RX J2052.2+4419, is detected at a mean count rate of 
0.015 cts/s, thus giving a total of 67 photons.

A second ROSAT HRI observation was performed on Dec. 2--8, 1997 in order to
follow-up on this surprising detection. \v7\ was again detected
at a mean rate of 0.010 cts/s during this 10.8 ksec observation, yielding
a total of 108 photons.

The area around \v7\ had been scanned during the ROSAT all-sky survey
in Nov. 1990 for a total of 370 sec. No X-ray emission is
detected from \v7\ during the PSPC scans, giving a 3$\sigma$ upper limit 
of 0.019 cts/s.

The location of \v7\ was serendipituously observed in a pointed PSPC 
observation of the supernova remnant G84.2--00.8 (PI: Pfeffermann)
on Nov. 11, 1992. Again, no X-ray emission is detected from \v7.
Though located at an offaxis angle of 49\amin\ in the ROSAT PSPC, 
the derived 3$\sigma$ upper 
limit of 0.0058 cts/s is more stringent than that of the all-sky survey
due to the larger exposure time. 

A summary of all X-ray observations with ROSAT of the \v7\ location is given
in Tab. \ref{xlog}. Note that the count rates refer to the specific detector,
and that the HRI is considerably less sensitive than the PSPC.

\subsection{Archival IUE observations}

While searching the available archival data on \v7\ we became aware of
unpublished IUE observations on April 25, 1985.
The object appears in the observing log as A2050+455, and its coordinates 
coincide within 2\asec\ with the position of \v7. 
We have checked the position of the
aperture, from the coordinates of the guide stars, and it coincides with
the position of V751 Cyg within a few arcsec.
The magnitude, as given by the Fine Error Sensor onboard IUE was 14.2 mag. 
The absence of any other closeby object and the coincidence of the measured 
magnitude with the AFOEV (Association Francaise des Observateurs d'Etoiles
Variables) database (m $\approx$ 13.8) confirm the identification. 
V751 Cyg was therefore at or close to optical maximum
at the time of the IUE observation.

The object was observed in low dispersion through the large aperture in both 
wavelength ranges: 1200-1950 \AA\ (65 min; SWP25774) and 1800-3300 \AA\ (47
min; LWP05819).

The spectra have been extracted from the bi-dimensional files 
with the ESA-INES system (Rodr\'{\i}guez--Pascual \etal\ 1998).

\subsection{Optical observations}

\subsubsection{Photometric observations}

\v7\ is monitored by a number of amateur astronomers around the world,
among them, one of us (LS). A collection of brightness estimates over
the last 8 yrs is plotted in Fig. \ref{lc} showing \v7\ to vary irregularly
between 13.6 and 14.5 mag most of the time, consistent with earlier 
observations (Wenzel 1963). However, starting somewhere between 
March 1 and March 11, 1997, \v7\ dropped in brightness to as low as 
$\approx$17.8 mag. Only recently, \v7\ has started to recover from this 
low state.

We obtained photometric observations on two occasions.
First, we observed \v7\ on September 30, 1997 (coincident with 
the end of our spectroscopic runs; see below) with the 
Danish photometer installed at the 1.5\,m telescope of the Observatorio 
Astron\'omico Nacional de San Pedro M\'artir, Mexico (OAN-SPM). It was set-up
to allow observations in the Str\"omgren photometric system.
Three measurements of \v7\ were taken together with standard stars.
In addition, for the time of the second HRI observation we performed 
multicolor photometry
with the 1.5\,m telescope of OAN-SPM to ensure broad-band coverage
in the Johnson UBVRI system. Standard fields PG2213-00.6 and RU 149 from 
Landolt (1992) were observed in the same night.

\subsubsection{Spectroscopic observations}

After the X-ray detection of \v7\ we obtained optical spectra on the nights
of September 28--30, 1997 at the 2.1\,m telescope of OAN-SPM 
(Figs. \ref{spec}, \ref{speccomp}).
The Boller \& Chivens spectrograph was used with a 300 l/mm  (400 l/mm in 
the last night) grating,
thus covering the 3600--7500 \AA\ (5500--8500 \AA) range. In combination 
with a 2\arcsec\, slit a   8 (6) \AA\ FWHM resolution is reached. 
Exposure times of 600 sec (1200 sec during last night) were chosen.
The spectrophotometric standard stars BD\,284211 and G191-B2B were observed 
with the same settings as \v7\ for flux calibration. During the last night the
slit was inclined to accommodate a nearby star for tracing the flux variations,
since the weather was not very good and tiny clouds were passing by.

The observational data were reduced using the IRAF package DAOPHOT and the
corresponding routines for long slit spectroscopy. The optimal extraction 
method was used for extraction of the spectra from the two-dimensional images.

\begin{table*}
\caption{List of X-ray sources found during the two HRI observations}
\vspace*{-0.3cm}
\footnotesize\hspace*{-0.75cm}
\begin{tabular}{cccccccc}
\hline
\noalign{\smallskip}
         & \multicolumn{2}{c}{HRI pointing 1} & 
           \multicolumn{2}{c}{HRI pointing 2} &  & &   \\
\hline
    Name & R.A.~~~~~~~Decl.    & CR$^{(1)}$  &
           R.A.~~~~~~~Decl.    & CR$^{(1)}$  & $\!\!$offaxis$\!\!$  & 
      D$^{(2)}$ & Ident. (Type) \\
         & (2000.0)     & $\!\!\!\!$(10$^{-3}$ cts/s)$\!\!$ & 
           (2000.0)     & $\!\!$(10$^{-3}$ cts/s)$\!\!\!$ & angle   &  
          (\asec)   &        \\ 
\noalign{\smallskip}
\hline
\noalign{\smallskip}
 RX J2052.1+4426 &                                      &  $<$0.76 &
                   20\H52\M09\fss2  +44\D26\amin04\asec &  1.5$\pm$0.4 &
   ~\,6\farcm9  & 4.3 & HD 198931 (B1V)  \\
 RX J2051.9+4425 &                                      &  $<$0.53 &
                   20\H51\M59\fss0  +44\D25\amin44\asec &  1.5$\pm$0.4 &
   ~\,6\farcm9  & 7.9 & LkHA 164 (EM*) \\
 RX J2053.2+4423 & 20\H53\M15\fss2  +44\D23\amin17\asec &  7.4$\pm$1.4 &
                   20\H53\M14\fss5  +44\D23\amin14\asec &  6.8$\pm$0.9 &
   12\farcm0  & 1.4 & HR 8001   (B5V)$^{(3)}$   \\
 RX J2053.9+4423 & 20\H53\M54\fss1  +44\D23\amin09\asec & 531$\pm$12 &
                   20\H53\M53\fss7  +44\D23\amin11\asec & 412$\pm$7 &
   18\farcm7  & 2.6 & HD 199178   (G2V)  \\
 RX J2050.8+4421 &                                      &  $<$1.59 &
                   20\H50\M51\fss5  +44\D21\amin47\asec &  2.4$\pm$0.8 &
   14\farcm6  &  &   \\
 RX J2052.2+4419 & 20\H52\M12\fss9  +44\D19\amin26\asec & 14.6$\pm$1.8 &
                   20\H52\M12\fss3  +44\D19\amin24\asec & 10.2$\pm$1.0 &
   ~\,0\farcm3  & 0.9 & $\!\!$\v7\ (VY Scl)$\!\!$    \\
 RX J2052.4+4417 &                                      &  $<$0.75 &
                   20\H52\M26\fss6  +44\D17\amin04\asec &  0.6$\pm$0.3 &
   ~\,3\farcm4  &  &   \\
 RX J2052.2+4415 & 20\H52\M13\fss3  +44\D15\amin30\asec &  1.1$\pm$0.5 &
                                                        &   $<$0.36      &
   ~\,3\farcm7  &     &      \\
 RX J2052.7+4414 & 20\H52\M44\fss2  +44\D14\amin35\asec &  1.5$\pm$0.6 &
                                                        &    $<$0.65    &
   ~\,7\farcm4  &     &     \\
 RX J2051.0+4411 & 20\H51\M04\fss2  +44\D11\amin34\asec &  3.6$\pm$1.2 &
                   20\H51\M02\fss8  +44\D11\amin39\asec &  3.2$\pm$0.8 &
   14\farcm3  &     &      \\
 RX J2053.4+4410 &                                      &   $<$1.79      &
                   20\H53\M25\fss1  +44\D10\amin32\asec &  3.0$\pm$0.9 &
   15\farcm7  &     &      \\
 RX J2051.4+4408 & 20\H51\M25\fss6  +44\D08\amin19\asec & 11.8$\pm$1.9 &
                   20\H51\M24\fss7  +44\D08\amin16\asec & 13.9$\pm$1.4 &
   13\farcm7  &4.6 & $\!\!$BD+43 3744$\!\!$   (G5)  \\
 RX J2052.9+4407 & 20\H52\M58\fss4  +44\D07\amin18\asec & 31.4$\pm$2.9 &
                   20\H52\M58\fss0  +44\D07\amin16\asec & 121$\pm$3.6 &
   14\farcm5  &     &       \\
 RX J2051.4+4404 & 20\H51\M27\fss3  +44\D04\amin26\asec &  7.2$\pm$1.7 &
                   20\H51\M28\fss1  +44\D04\amin11\asec &  3.8$\pm$1.2 &
   16\farcm8  &  &      \\
\hline
\noalign{\smallskip}
\end{tabular}

\noindent{\Ni\small 
   $^{(1)}$ Upper limits are 3\,$\sigma$ confidence level. \\
   $^{(2)}$ Distance between best-fit X-ray position and optical position of
       presumed counterpart (the better of the two X-ray positions).\\
   $^{(3)}$ The measured flux is consistent with being no
       X-ray emission, but solely UV emission of this $m_{\rm B}$=4.6 mag star
       due to the UV leak of the HRI (Bergh\"ofer 1997).
     }
\label{xlist}
\end{table*}

\section{Results}

   \begin{figure}
    \vbox{\psfig{figure=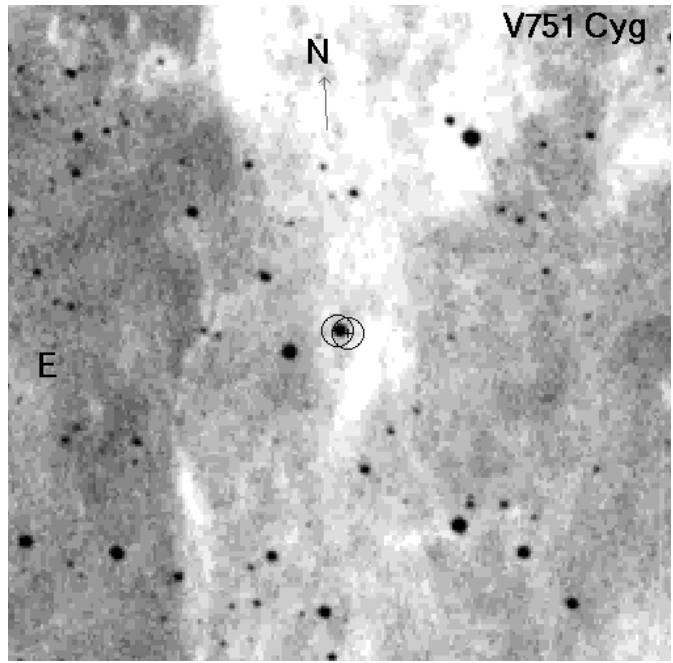,width=8.8cm,%
          bbllx=2.6cm,bblly=10.8cm,bburx=18.3cm,bbury=26.4cm,clip=}}\par
    \caption[fc]{A 7\amin$\times$7\amin\ digitized sky survey image with 
           the 10\asec\ X-ray error
           radii of the two HRI observations overplotted. V751 Cyg is the only 
           visible optical source down to $\approx$20\m. The grey colors 
           are due to the diffuse, nebular optical emission in this region.  }
    \vspace{-0.1cm}
      \label{fc}
   \end{figure}

\subsection{Identification of the X-ray source with \v7}

In order to ensure a correct identification of the detected X-ray emission
with \v7, we have investigated several alternatives:
(1) Check of the attitude solution: There are no indications of any anomaly in 
the guide star acquisition. Next, from the 14 X-ray sources in the HRI field of
view, five sources in addition to \v7\ have rather secure optical 
identifications based on the positional coincidence, spectral type and 
the $L_{\rm X}/L_{\rm opt}$
ratio. Also, a few sources have been detected during the
all-sky survey with positions which are consistent within their errors
to those derived from our pointings, i.e. \#4 $\equiv$ 1RXS J205353.7+442308, 
\#12 (detected with a count rate of
0.018$\pm$0.07 cts/s (just below the count rate threshold for inclusion in the
1RXS catalog) at R.A. = 20\h51\m28\fss2, Decl. = 44\degr 08\amin 02\asec)
 and \#13 $\equiv$ 1RXS J205257.8+440716.
Finally, 9 out of the 14 sources are detected in both of the two independent
pointings, including \v7.
(2) Mis-identification: Given the correct attitude solution and the fact that
there is no optical object brighter than about 20th mag (due to some nebulosity
associated with the Cyg T1 association) within the 10\asec\ X-ray error box
(see Fig. \ref{fc}),
we consider a mis-identification as very improbable. 
(3) Background source: 
Due to the very soft X-ray spectrum
a variable X-ray source located along the line of sight towards \v7
must have a distance smaller than the molecular clouds at 800--920 pc. 
Such a source, if it were not \v7, cannot be excluded, but its peculiar 
properties (X-ray intensity amplitude; no accompanied optical variability) 
are unlike any known variable X-ray source population.

\subsection{Relation of optical and X-ray intensity}

Combining the optical monitoring data with the X-ray upper limits and detection
of \v7\ (Fig. \ref{lc}) reveals that the two X-ray non-detections occurred
during the optical high state whereas the X-ray detections occurred during the
optical low state. We therefore find  evidence supporting an 
anti-correlation of optical and X-ray intensity in \v7.

\subsection{The extinction towards and distance of \v7}

   \begin{figure}
    \vbox{\psfig{figure=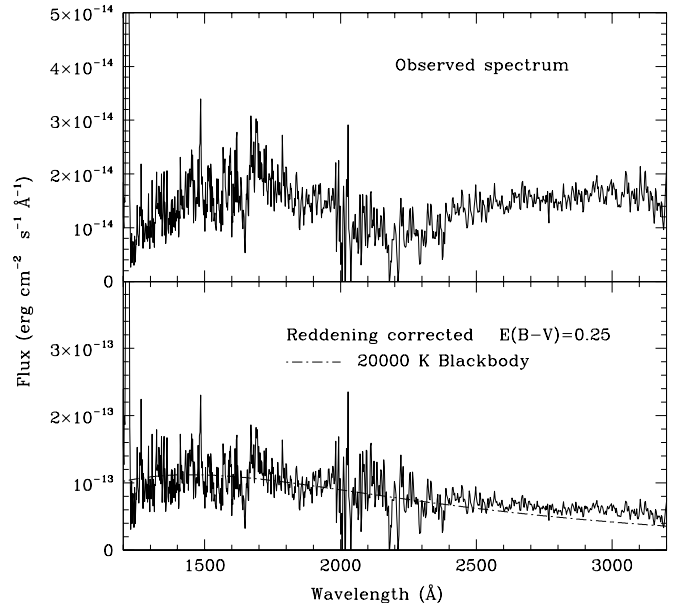,width=8.8cm,%
          bbllx=0.8cm,bblly=6.8cm,bburx=18.1cm,bbury=22.7cm,clip=}}\par
    \caption[nuf]{Archival, observed IUE spectrum of \v7\ (top) taken on 
            April 25, 1985. The two peaks  near 1480 and 2020 \AA\ are cosmic 
           rays affecting the spectrum. 
           The spectrum corrected for extinction of $E_{\rm B-V}$=0.25
           is shown in the bottom panel together with a blackbody model
           of 20000 K.
          }
      \label{iue}
   \end{figure}

The UV spectrum shows a broad absorption centered at  2200 \AA\, which is the
typical signature of interstellar absorption. The reddening toward V751 Cyg
has been estimated by removing this feature by using a standard galactic
interstellar absorption law and different values of the color excess. The
best result is obtained with $E(B-V)=0.25\pm0.05$. No assumption
about the intrinsic spectral shape of the spectrum has been made to
evaluate the correction.

With the intrinsic color $(B-V)$ being near zero, the above color excess
implies a visual extinction of $A_{\rm V} = 0.82\pm0.17$. Using a mean 
extinction law of 1.9 mag/kpc (Allen 1973) we derive
a distance to \v7\ of 430$\pm$100 pc (the error being solely due to the
error in $E(B-V)$). Alternatively, using the extinction
distribution as estimated by Neckel \& Klare (1980) results in d 
$\approx$ 610$\pm$30 pc. Because of the global nature of the extinction 
determination in both cases and the possibility that circumbinary extinction
cannot be excluded (which would reduce the derived distance), 
we will adopt d=500 pc in the following.
We mention that this is consistent with the previous
upper limit of 800 pc (Dobashi \etal\ 1994).
For the sake of ease in the later sections (describing the X-ray spectral fits)
we note that the above 
$A_{\rm V}$ corresponds to a neutral hydrogen column density of 
$N_{\rm H}$ = (1.1$\pm$0.2)$\times$10$^{21}$ cm$^{-2}$, using 
the relation $A_{\rm V}$ = 17/23$\times$ $N_{\rm H}$ [10$^{21}$ cm$^{-2}$]
as derived by Predehl \& Schmitt (1995).
The above values and a maximum and minimum brightness of 
$m_{\rm V}^{max} = 13.2$ mag (Martynov 1958) and 
$m_{\rm V}^{min} = 17.8$ mag (Ouda team CCD photometry, from VSNET) imply
absolute magnitudes of
$M_{\rm V}^{max} = 3.9$ mag and $M_{\rm V}^{min} = 8.5$ mag.

Even after the reddening correction, we do not see a rising continuum
toward the shorter wavelengths which could be attributed to the presence 
of a hot white dwarf in this binary system.

The UV spectrum (Fig. \ref{iue}) has a low S/N and it is nearly featureless, 
except for a prominent absorption near 1640 \AA, which can be ascribed to HeII.
Other features, as the emissions near 1590 \AA\ and 3200 \AA, are cosmic rays
affecting the spectrum. No other emission line can be unambiguously identified.
In particular, there is not any noticeable feature near the positions of the 
\ion{N}{V} 1243 \AA, \ion{Si}{IV} 1400 \AA, \ion{C}{IV} 1550 \AA,  or Mg\II\ 
resonance lines which are seen in other VY Scl 
stars during high state (see e.g. LX Ser; Szkody 1981, 
MV Lyr, Szkody \& Downes 1982, or TT Ari, 
Shafter \etal\ 1985; PX And, Thorstensen \etal\ 1991; BH Lyn, 
Hoard \& Szkody 1997). 
UV spectra of VY Scl stars at different brightness levels are
shown in la Dous (1993). In particular, the spectrum described here resembles 
most that of TT Ari in high state, although the spectral features are less
marked in the case of V751 Cyg.

\subsection{The inclination of the \v7\ binary}

La Dous (1991) has shown in her study of a large number of IUE
spectra of dwarf novae and nova-like stars, that there exists a clear
relation between the inclination of the system and the equivalent width of
some UV lines.  Her sample includes nine VY Scl stars. 
The VY Scl sample follows the same trend as other nova-like stars: for
high values of inclination ($\gax$70\degr), the lines are
in emission, with large equivalent widths, and for lower inclinations the
equivalent widths decrease so that for nearly face-on systems the lines
are in absorption.  

The lack of evident emission lines in the spectrum of V751 Cyg (Fig. \ref{iue})
therefore points to a low inclination. Although the S/N ratio of the IUE 
spectrum is low, we have estimated the equivalent widths of some 
lines, namely \ion{Si}{IV} 1400\AA\ (EW = 2.7 \AA), 
\ion{C}{IV} 1550\AA\ (13 \AA), 
the blend \ion{Al}{III}+\ion{Fe}{III} 1860\AA\ (5.5 \AA) and 
\ion{Mg}{II} 2800\AA\ (0.7 \AA)
(note that in la Dous' notation positive equivalent widths
mean absorption lines). A comparison with the equivalent widths of the VY Scl 
systems with known
inclination (taken from the compilation of Greiner 1998), results in a value 
for the orbital inclination of V751 Cyg of $i$=30$\pm$20\degr.
In any case, it can be said conservatively that the inclination of the
system is $<$50\degr.

\subsection{The X-ray spectrum and luminosity of \v7}

Given the sensitivity dependence of the HRI to the X-ray spectral shape
(soft vs. hard spectrum), the comparison of the PSPC and HRI 
count rates/upper limits requires some knowledge of the X-ray spectrum of
\v7\ during the HRI observation. Before discussing this below, we note that 
the conversion factors for the two extremes are 2.7:1 and 7.8:1 
(Greiner \etal\ 1996) for a very hard and a very soft spectrum, respectively. 
Thus, the HRI count rate of 0.0146 cts/s translates into a corresponding
PSPC count rate between 0.039 and 0.11 cts/s. 

   \begin{figure}
    \vbox{\psfig{figure=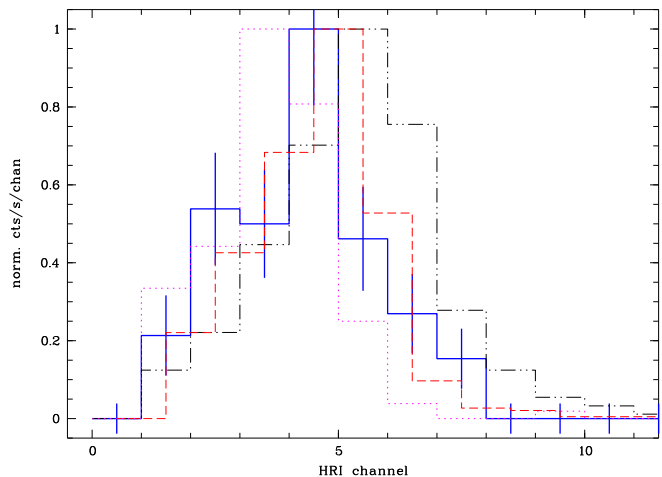,width=8.8cm,%
          bbllx=2.1cm,bblly=1.2cm,bburx=17.2cm,bbury=12.2cm,clip=}}\par
    \vspace{-0.25cm}
    \caption[pha]{Comparison of the HRI pulse height distribution of V751 Cyg
         (thick line) with those of an extremely absorbed hard X-ray source 
         (GRS 1915+105 - dash-dot-dot line), an moderately absorbed 
         supersoft source (RX J0925.7--4758 - dashed line)
         and a slightly absorbed supersoft source (AG Dra - dotted line).
         All sources have been observed on-axis, and the temporal gain
         change between the observations (distributed over three years)
         has been taken into account (corrected to the June 1997 observation 
         of \v7).   }
      \label{pha}
      \vspace{-0.15cm}
   \end{figure}

The need to quantify the X-ray spectral shape  has motivated us
to study the pulse height distribution in more detail. The CsI-coated
micro-channel plates of the HRI provides some crude spectral sensitivity,
and we have used two approaches to determine the spectral characteristics.
First, we have compared the pulse height distribution
of the \v7\ observation with other, on-axis observations of sources with
known (from PSPC observations at about the same time) soft X-ray spectra in 
order to derive a rough idea for the interpretation of the shape of the 
pulse height distribution of \v7.
Based on the description of the temporal gain changes of the HRI
as described in the recent HRI calibration report (David \etal\
1997), in particular their Fig. 20, we have then shifted the pulse height
distributions of these selected observations to match the gain status
of our HRI observation on June 3, 1997. The result is shown in Fig. \ref{pha}
and suggests that the \v7\ spectrum is neither an unabsorbed supersoft 
spectrum nor a strongly absorbed hard spectrum, but similar to \0927,
i.e. a soft source suffering some absorption.

   \begin{figure}
    \vbox{\psfig{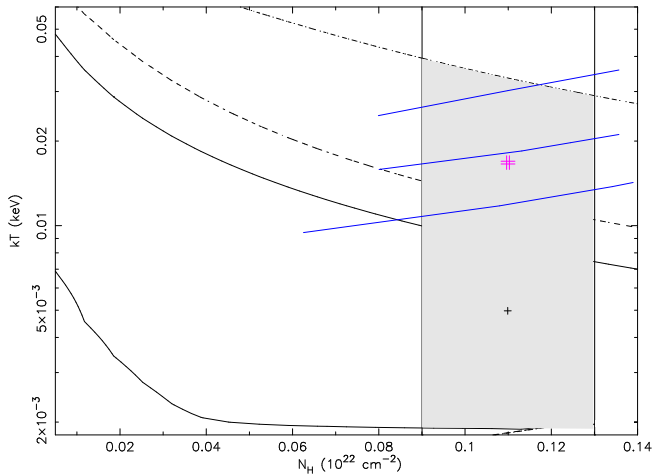}}\par
       \vspace*{-0.2cm}
    \caption[pha]{Spectral fit result of the June 1997 ROSAT observation
         of V751 Cyg:
        Confidence contours of the blackbody fit in the
         $kT-N_{\rm H}$ plane: 68\% (solid line),
         90\% (long-dashed line), 99\% (short-dashed line). 
         The two vertical lines denote the $N_{\rm H}$ range allowed
         by the IUE data ((1.1$\pm$0.2)$\times$10$^{21}$ cm$^{-2}$), 
        and the vertical extent of the hatched region marks the 99\% confidence
         region of the blackbody temperature.
         The three solid  lines crossing the hatched region
        mark contours of constant luminosity (at an assumed distance of 500 pc)
         of 10$^{34}$, 10$^{36}$ and 10$^{38}$ erg/s (from top to bottom),
         respectively.
   The cross denotes the best fit-value of $kT$=5\,eV and the double cross the
   parameter pair used in the following discussion (see text).
         }
      \label{ufs}
      \vspace{-0.2cm}
   \end{figure}

In a second step, we make use of new software developed to allow improved 
spectral analysis of HRI data (Prestwich \etal\ 1998).
The CsI coating on the microchannel plates means that the HRI has two-color 
energy resolution above and below 0.62 keV. We have used this spectral response
to  study the pulse height distribution of V751 Cyg in more detail.  Full 
details of the HRI spectral calibration will be given in a separate paper 
(Prestwich \etal\ 1998); here we give a summary of the 
calibration procedure.
A spatial gain map  and spectral response matrix (for the center of the 
detector) were produced from ground calibration data.  However, this response 
matrix cannot be used for in-flight data because the gain changes slowly with 
time throughout the mission, with occasional ``jumps'' when the gain is 
adjusted from the ground. To solve these problems, the spatial and temporal 
gain variations have been tracked using the Bright Earth (BE) data. The BE 
data is dominated by scattered solar X-rays: 
the dominant feature is the oxygen K$\alpha$ line at 525 eV.
It is effectively a monochromatic flat field, and is used to monitor gain 
changes throughout the mission.  Using the BE data and other calibration 
sources it is possible to shift the response matrix derived from ground-based 
observations to the gain state of any given observation. There is, however, 
one further complication: the HRI is ``wobbled'' and as the source is moved 
across the detector the gain may change.  Hence it is important to derive a 
response matrix weighted by livetime for a given source position. This is done 
by de-applying the aspect solution to
calculate the detector position and gain of a source versus time.

The observation of V751 Cyg on June 3, 1997 is on axis and was obtained  very 
shortly after the HRI gain was increased on May 13 1997.  Analysis of  BE data 
obtained after the change indicate that the maximum of the BE gain function 
lies at 
HRI PHA channel 4.5, requiring a shift in the ground-based response matrix of 
approximately 0.6 channels.  Fits using this response matrix to the photons 
extracted from a 15-pixel radius around the position of V751 Cyg and
background subtracted from a circular ring region of 8 pixel width result
in the confidence contour map shown 
in Fig. \ref{ufs}.  From this figure it is clear that simple black-body 
models with kT of a few tens of eV are consistent with the data, whereas 
higher temperature models (0.5 keV) can be ruled out.

Since the absorbing column cannot be constrained from the X-ray data, we fix
the absorbing column at $N_{\rm H}$ = 1.1$\times$10$^{21}$ cm$^{-2}$
as derived from the IUE spectrum (section 3.3). The best-fit temperature
of the blackbody model is 5 eV, and the formal 90\% confidence error
is about $^{+10}_{-5}$ eV (see the confidence contours in Fig. \ref{ufs}). 

The luminosity determination of \v7\ is a delicate problem because of
the softness of the emission (we measure just the very end of the Wien tail),
the uncertainties given by the detector response at these very soft 
energies, and the intercorrelation of spectral shape and absorbing column.
In addition, the deduced luminosity also depends on the applied model
with white dwarf atmosphere models giving typically a factor of 10--100
lower luminosity than a blackbody model. Though for simplicity we applied
a blackbody model (also justified by the poor spectral energy resolution)
we took several steps in order to determine a minimal luminosity.
As a first step, we minimized the effect of the absorbing column
by fitting a gaussian line with a width that
corresponds to the energy resolution of the detector and
without applying any absorbing column. This results in a best-fit centroid 
of 9 eV and a luminosity of 2$\times$10$^{31}$ erg/s in the 0.1--2.4 keV range.
As the next step, we introduce absorption, i.e. fit an absorbed gaussian line
with the absorbing column fixed at an amount as derived from the IUE data.
The result is a centroid energy of 2 eV (at the fixed 
$N_{\rm H}$=1.1$\times$10$^{21}$ cm$^{-1}$),
and the luminosity in the ROSAT band (0.1--2.4 keV) alone is 
5$\times$10$^{33}$ (D/500 pc)$^2$ erg/s. Considering the full flux under 
the gaussian, e.g. adding up the flux below 0.1 keV, rises the luminosity to 
7$\times$10$^{33}$ (D/500 pc)$^2$ erg/s. We consider this the absolutely 
minimal flux which can be accomplished
by the IUE-derived extinction and the HRI detection, since any other model
will have a wider shape than a gaussian function.
Since the best-fit value derived with the blackbody model implies
extremely high luminosity, we adopt $kT = 15^{+15}_{-10}$ eV in the following
(and $N_{\rm H}$=1.1$\times$10$^{21}$ cm$^{-1}$ from the IUE spectrum). 
At this temperature the bolometric, unabsorbed luminosity is
$L=6.5\times 10^{36}$ (D/500 pc)$^2$ erg/s.

The HRI pointing on Dec. 3, 1997 shows \v7\ to be in the same intensity and 
spectral state as during the June 1997 observation. This suggests that
the soft X-ray emission is present throughout the whole optical low state.

\subsection{Re-assessment of the HEAO-1 detection of \v7}

\v7\ has been listed as the only detected VY Scl star besides MV Lyr
within the HEAO-1 A2 survey of more than 200 cataclysmic variables
(Cordova \etal\ 1981). This survey used data of one of the low-energy
detectors (LED1), a proportional counter sensitive in the
0.18--2.8 keV range (Rothschild \etal\ 1979). The field of view was
collimated to 1\fdg5 (FWHM) in and 3\grad\ (FWHM) perpendicular to the
scan direction. \v7\ was scanned during  Nov.\,25--Dec.\,2, 
1977 and is reported with an upper limit of $<$2.56 cts/sec at
0.18--0.48 keV and a detection at 3.95$\pm$0.90 cts/sec in the
0.48--2.8 keV band (Cordova \etal\ 1981). Using a thermal bremsstrahlung
spectrum with $kT=10$ keV and $N_{\rm H}$ = 1$\times$10$^{20}$ cm$^{-2}$
Cordova \etal\ (1981) give a conversion rate of 
3$\times$10$^{-11}$ erg/cm$^2$/s per LED1 cts/sec in the 0.48--2.8 keV band. 
Though the absorbing column to \v7\ is larger than this assumption, the effect 
on the absorption corrected flux is not large. Thus, the detected count rate
corresponds to 1.2$\times$10$^{-10}$ erg/cm$^2$/s in the 0.48--2.8 keV band.

   \begin{figure}
    \vbox{\psfig{figure=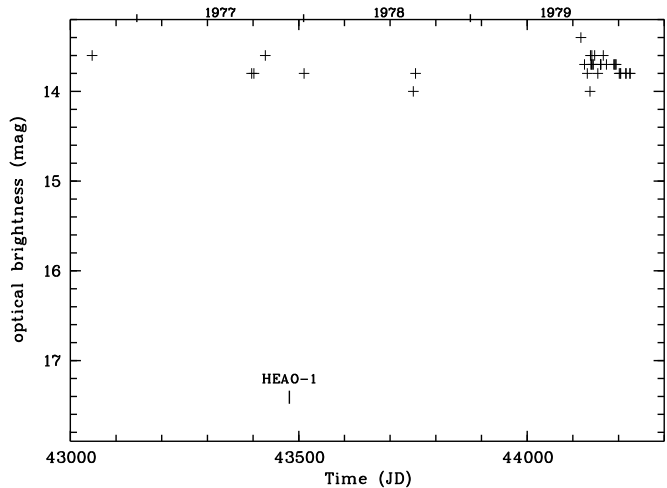,width=8.8cm,%
           bbllx=2.4cm,bblly=15.9cm,bburx=18.0cm,bbury=27.5cm,clip=}}\par
    \caption[pha]{Optical light curve of V751 Cyg (data from AAVSO) 
            around the time of the
            HEAO-1 A2 observation in 1977. Though the coverage is sparse, it
            seems justified to assume that V751 Cyg was in its optical
            high (normal) state during that X-ray observation.
           }
      \label{heao}
   \end{figure}

This flux corresponds to a ROSAT PSPC count rate of 16 cts/s, much higher than
any of our ROSAT measurements of \v7. The reported HEAO-1 A2 count rate is 
also a factor of 5 larger than that of MV Lyr (Cordova \etal\ 1981), the 
second brightest VY Scl star at X-rays (Greiner 1998). This 
motivated us to check the optical state of
\v7\ during the HEAO-1 A2 observation as well as to investigate
the possibility of mis-identification of the X-ray emission detected with
HEAO-1 A2.

Data from the AAVSO database (courtesy J. Mattei) indicate (though the
coverage is poor) that \v7\ was seemingly in the optical high state during 
the HEAO-1 A2 observation.
Given the same optical state during the HEAO-1 A2 and the ROSAT PSPC 
observations but the huge difference
in X-ray intensity, it is extremely unlikely that the identification of the
HEAO-1 A2 X-ray source (with a position uncertainty of 1\fdg5--3\degr)
with \v7\ is correct.

   \begin{figure}
    \vbox{\psfig{figure=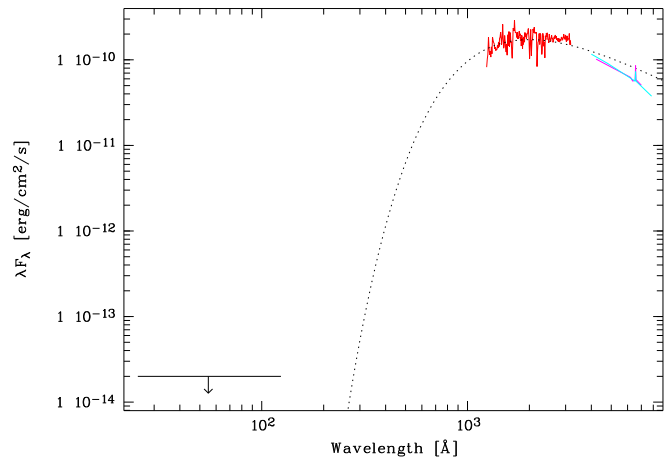,width=8.8cm,%
          bbllx=1.4cm,bblly=1.2cm,bburx=17.2cm,bbury=12.2cm,clip=}}\par
    \caption[nuf]{Broad-band energy spectral distribution of V751 Cyg
           during the optical high-state. Combined are observations of 
           different epochs, but all during optical high-state: 
           The ROSAT upper limit from Nov. 1992,
           the IUE spectrum from April 1985,
           the optical spectra (both at $m_{\rm V}=14.3$) from May 1986 
            (Downes \etal\ 1995) and Oct. 1994 (Munari \etal\ 1997).
           Overplotted is a 3$\times$10$^4$ K accretion disk spectrum 
           demonstrating that the UV/optical emission can be well reproduced
           by optically thick emission.}
      \label{nufnu}
   \end{figure}

We therefore have attempted to identify a better candidate counterpart
for the HEAO-1 A2  X-ray source.
The ROSAT all-sky survey data reveal more than a dozen X-ray sources
within 3\degr\ of \v7, the brightest of which is the
FK Com star (spectral type G2V) HD 199178 (Bopp \& Stencel 1981). 
However, it has a quiescent count rate of 1.6 cts/sec, a factor
of 10 too low to account for the emission seen with HEAO-1 A2. Even adding
up the fluxes of all ROSAT sources which potentially could have been 
within the collimated LED1 field of view  would give about 4 cts/s, 
still a factor 4 below the HEAO-1 A2 rate.
On the other hand, HD 199178 was around maximum of its 9 yr activity
cycle at the end of 1977 (Jetsu \etal\ 1990), and it is thus conceivable that
it was observed with HEAO-1 A2 during a flare. While this possibility can 
be checked by a temporal analysis of the HEAO-1 A2 data, we
propose tentatively that HD 199178 (which is only 18\farcm5 away from \v7)
is a better counterpart candidate than \v7\ to 
the X-ray source detected with HEAO-1 A2.

\subsection{The optical emission}

The Str\"omgren photometry performed on September 30, 1997 yields the 
following results for \v7: v=16.28$\pm$0.06, b-y=0.17$\pm$0.02,
m1=0.16$\pm$0.02, c1=--0.14$\pm$0.03, $\beta$=3.95$\pm$0.02
(including systematic uncertainties).
The narrow Str\"omgren  u, b and y filters measure mostly continuum 
in cataclysmic objects, although HeII 4686 \AA\ contributes in the b band. 
The v and $\beta$ filters are dominated by H$\gamma$ and H$\beta$, 
respectively. Thus, the c$_1$
index c$_1$=(u-v)-(v-b) is a reflection of the Balmer discontinuity
(it is negative, when the Balmer jump is in emission), while m$_1$=
(v-b)-(b-y)  is a more complex index.
For the possible interpretation of the above mentioned magnitudes we 
refer to the survey of dwarf novae by Echevarria \etal\ (1993).
The comparison of the  obtained values shows that \v7\ at the moment of the
observation was occupying an extreme place in 
the range presented  by  dwarf novae, but with some reservations could be 
fitted in between quiescent systems and systems in rise, which actually 
corresponds to the optical state of \v7\ during this time.  
The most extreme value is that of $\beta$ which is 
much higher than most of the systems caught in the maximum or rise. This 
could partially be due to the contribution of the nebular background emission,
but it also is consistent with a reduced mass transfer rate and low intensity 
contribution of the disk.

The broad band photometry on Dec. 3, 1997 has found V751 Cyg at 
U=15.53$\pm$0.02, B=16.17$\pm$0.01, V=16.08$\pm$0.01, R=15.58$\pm$0.01, 
I=15.36$\pm$0.01. These errors are according to the accuracy of the 
aperture photometry and do not include systematic uncertainties 
(see Fig. \ref{speccomp})  and the magnitudes are 
corrected for the air mass by interpolation of the  observed standard fields. 
However, extinction coefficients were not calculated in accordance to the
generally adopted technique, so in Fig. \ref{speccomp}, the error bars reflect 
possible deviations of magnitudes due to the extinction. In general, there is 
good agreement between the September spectrophotometry and the December broad 
band photometry which
was conducted simultaneously with the second HRI observation.

\begin{figure}
    \vbox{\psfig{figure=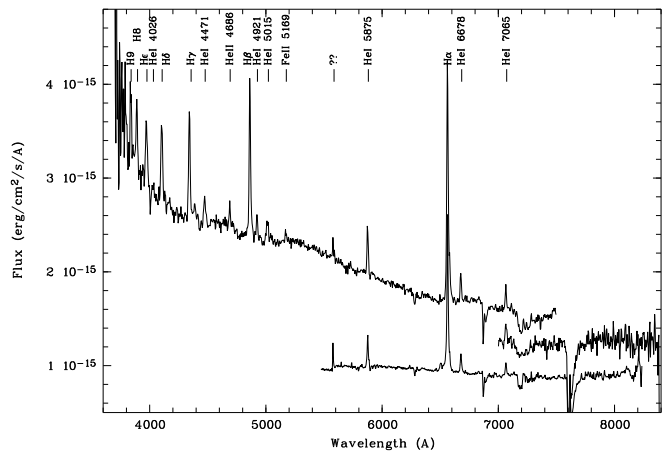,width=8.8cm,%
          bbllx=1.2cm,bblly=1.cm,bburx=17.2cm,bbury=12.2cm,clip=}}\par
    \caption[pha]{Composite of a blue- and red-arm spectrum of V751 Cyg 
         taken on Sep. 30, 1997 (JD = 2450722). The continuums intensity 
         around 5500 \AA\ corresponds to $m_{\rm V} \approx$ 15.4 mag. 
         The spectra have not been reddening corrected.}
      \label{spec}
   \end{figure}

\begin{figure}
    \vbox{\psfig{figure=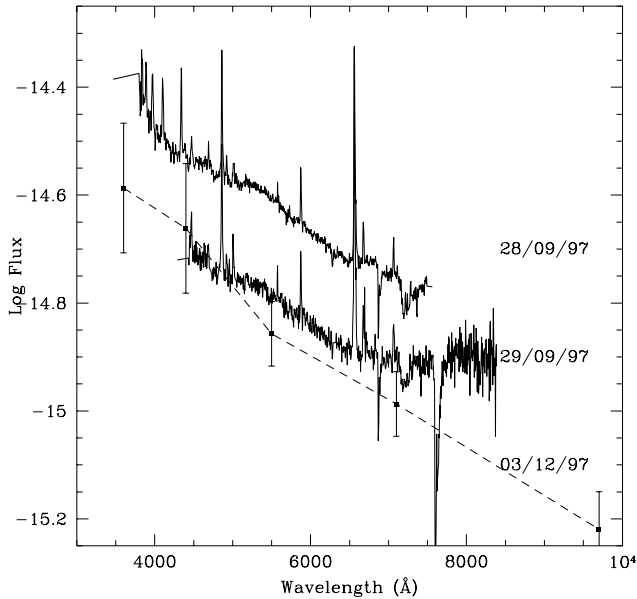,width=8.5cm,%
          bbllx=0.5cm,bblly=5.6cm,bburx=20.4cm,bbury=24.3cm,clip=}}\par
    \vspace{-0.2cm}
    \caption[pha]{Comparison of the variability of the global shape
         of the optical spectrum within a few days.
         For Dec. 3, 1997 (the time of the second ROSAT HRI observation)
         we have connected our UBVRI photometry by a dashed 
         line to visualize the global shape. The error bars reflect
         both the errors of transformation and extinction uncertainty.}
      \label{speccomp}
   \end{figure}

In the September 1997 spectroscopy
we detect strong and relatively narrow emission lines of the Balmer series up 
to H9, and 
many strong HeI emission lines (see a sample spectrum in Fig. \ref{spec}). 
The high-excitation He\,{\sc II} 4686 \AA\ 
line is moderately strong while C\,{\sc III}/N\,{\sc III} 4640--4650 \AA\ 
is present but very weak. The Balmer decrement is very shallow. The mean ratio 
of H$\delta$:H$\gamma$:H$\beta$:H$\alpha$ is 0.65:0.8:1.0:1.6, thus implying
a rather high electron density and temperature.
According to Drake \& Ulrich (1980) the electron densities $N_{\rm e}$ 
would range from $10^{13}-10^{14}$ cm$^{-3}$ with 
relatively low optical depth  in L$\alpha$ of the order of $10^3-10^4$
and a relatively weak ground-state photoionization rate of 
$3\times10^{-2}$ s$^{-1}$. The Balmer decrement is less sensitive to 
the effective temperature and fits to the narrow  range 
of temperatures $(2-4)\times10^4$ considered by Drake \& Ulrich (1980).
The FWHM of the H$\alpha$ line varies between 12 and 17 \AA\ along with 
significant variations of the continuum level and slope of the spectrum.
Since we used a narrow slit (but did not orient it along the 
parallactic angle), and atmospheric conditions were far from 
perfect, we calibrated the last night observations by a field star
which was placed in the slit. For \v7\ we detected significant variations of 
continuum intensity and slope of the spectra, especially
from one night to the other which hardly can be accounted for imperfection of 
our observing/reduction techniques and probably are due to the intrinsic 
variations of the \v7. However,
we do not find continuous variations of intensity
or radial velocity with the suspected orbital period of 6 hrs (though
the observations covered only 4 hrs).
Extreme cases of spectra at different nights are shown in Fig. \ref{spec}.

The emission line spectrum 
could be interpreted as arising from the wind of the secondary
or alternatively as an optically thin accretion disk.
The lines are not very broad (half widths at the base of
$\sim$1200 km/s), probably due to orbital and/or
turbulent motion in the low inclination disk. Also, the high-excitation line 
He\,{\sc II} 4686 \AA\ is clearly detected though  not as strong as in 
magnetic cataclysmic variables or supersoft X-ray binaries. Finally, the
shallow Balmer decrement of \v7\ is  similar (though somewhat steeper) 
to that observed in some other VY Scl stars at intermediate brightness levels 
(e.g. in Sep. 1980 in MV Lyr; Fig. 5 in Robinson \etal\ 1981). In contrast, 
the VY Scl star low state spectra have very narrow and fainter emission lines 
which have been interpreted
to arise in the heated chromosphere of the secondary. Based on these
three facts and a $m_{\rm V}$$\approx$15.4 (i.e. at least 2 mag brighter
than during the low-state) deduced from a line-free region
of our spectra around 5500 \AA\ we conclude that
\v7\ was observed in an intermediate state. It is, however, difficult 
to assess whether the accretion disk disappeared in the early phase of 
the low state and had been reestablished by September
or whether a residual disk always remained.

The detection of the high-excitation \ion{He}{II} 4686 \AA\ line during our
intermediate state spectroscopic observations in September 1997 is in line 
with the X-ray detections in June and December 1997, since its ionization
implies the presence of E$>$54 eV photons. This confirms the conclusion drawn
from the December 1997 HRI observation that seemingly the soft, luminous 
X-ray emission is ``on'' throughout the optical low state and not just a 
short-lived, temporary event. 
We therefore may anticipate that the He\,{\sc II} emission
is even stronger during the low state, and that its intensity declines
together with that of the X-ray emission as the optical brightness rises
towards maximum.

We do not detect any signature of the companion in the red part of the 
spectrum. This is compatible with the expectation of a M0-1 main-sequence star
according to a $P\approx$ 6 hrs binary and the mass-radius relation of 
Patterson (1984) which would have $m_{\rm V}\approx 18$ mag at 500 pc. 
We note that an early- to mid-F type companion as suggested by 
Downes \etal\ (1993) is far too bright to be consistent with our spectra and 
$d \lax 1$ kpc.

\section{V751 Cyg}

\subsection{V751 Cyg as a supersoft X-ray binary}

The following picture emerges.
During optical low states \v7\ exhibits transient soft X-ray emission 
thus revealing itself as a  supersoft X-ray binary.
The appearance of \ion{He}{ii} 4686 \AA\ emission in optical spectra taken 
nearly simultaneous with the ROSAT HRI data also indicates the presence of
photons with energy $>$54 eV.
V751 Cyg, like the other members of the VY Scl star group,
accretes at a few times 10$^{-8} M_\odot$ yr$^{-1}$. If the mass of the
white dwarf in \v7\ is small, this may allow nuclear burning as the
high X-ray luminosity suggests. It is worth emphasizing that recent 
calculations of hydrogen-accreting carbon-oxygen 
white dwarfs have shown that the accretion rate for low mass white dwarfs
(0.5--0.6 \msun) can be as low as 1--3$\times$10$^{-8}$ \msun/yr 
(Sion \& Starrfield 1994, Cassisi \etal\ 1998) while still maintaining 
shell burning (consistent with Fujimoto 1982).
The \v7\ values of $M_{\rm V}^{\rm max} = 3.9$ (see section 3.3) and 
${\log \Sigma} = \log (L_x/L_{\rm Edd})^{1/2} P_{\rm orb}^{2/3} (hr) = -0.23$ 
are consistent, within the uncertainties of $L_x$ and $P_{\rm orb},$ with the 
relation $M_{\rm V} = 0.83(\pm 0.25) - 3.46 (\pm 0.56) \log \Sigma$
found for 5 SSB (van Teeseling \etal\ 1997) implying
that, if nuclear burning is the correct interpretation
of the X-ray flux during the optical low state,
then nuclear burning may continue during the optical high state.

The explanation for the character of the optical and UV observations 
is not yet clear, but it seems certain that the 
illumination of the donor and disk
play important roles in determining what we see.  
If the X-ray source during the optical low state indeed is very luminous
one may expect a strong heating effect on the secondary as well as on
the accretion disk. 
The heating of the secondary in \v7\ is probably comparable to
that in supersoft X-ray binaries because the illumination depends on the
ratio of companion radius and binary separation which is similar in both
kind of systems. In addition, second order effects are competing against each
other, e.g the mass ratio dependence of the illumination would imply larger
heating in supersoft X-ray binaries while the flared disks in these systems
are also thought to occult the donor near the equatorial plane from 
illumination.
 Unfortunately, no photometry has been
obtained during the optical low state to immediately test for this effect 
in \v7\, though it is anyway not expected to produce a strong modulation 
due to the low inclination (see section 3.3).

The question of the illumination of the accretion disk has to be addressed
separately for optical low and high state.
As mentioned above, there is evidence in some VY Scl stars that during the
optical low state the accretion disk has vanished. Though we have no direct
evidence for this in \v7\ due to the lack of optical observations, 
the disk is certainly optically thin thus
drastically reducing the effectivity of illumination.
In the optical high state the illumination depends on 
whether hydrogen burning stops or whether it continues on 
an inflated white dwarf at a temperature below the sensitivity range of ROSAT: 
If the  burning stops then there are no soft X-rays which 
could be reprocessed. If the nuclear burning continues, 
reprocessing may still not be strong because the amount of reprocessing 
depends on the flaring of the accretion disk.
As was first shown by King (1997) and
later argued by Knigge \& Livio (1998), reprocessing of the radiation from the
white dwarf will begin to have a dominant effect on the local disk
temperature if the white dwarf luminosity
$L_{\rm WD} \gax 2.5 L_{\rm acc} (1-\beta)^{-1}$ (where $\beta$ is the albedo
of the disk surface). That is, a 
disk around a 1 \msun\ white dwarf accreting at 10$^{-8}$ \msun/yr 
will be dominated by reprocessing only if the white dwarf temperature is 
$>$2$\times$10$^5$ K $\equiv$ 17 eV.
This is seemingly just a value between the 
temperatures of SSS (30--50 eV) and V751 Cyg (15 eV), implying that one 
difference of the systems could be that the disk in V751 Cyg (if burning
continues during the optical high state) is not flared and therefore not 
dominated by reprocessing while the SSS disks are flared and
dominated by reprocessing and thus are optically much brighter than 
the VY Scl disks.

\subsection{Comparison of \v7\ to \rxj0513}

Our discovery of \v7\ as a transient supersoft X-ray source arose from the
similarity in the optical light curve of RX J0513.9--6951 and VY Scl stars.
In this section we discuss the similarities and differences between these
two systems. We note that, even if the underlying mechanisms are completely
different, the somewhat similar anticorrelations between X-ray and optical
have proved useful.

\rxj0513 shows $\sim$4 week optical low states which are accompanied
by luminous supersoft X-ray emission. It is generally assumed that the 
white dwarf accretes at a rate slightly higher than the burning rate, 
and thus is in an inflated state during the optical high state. 
Changes in the irradiation of the disk caused by the expanding/contracting 
envelope around the white dwarf have been proposed as explanation of the
1 mag intensity variation in \rxj0513\ (Southwell \etal\ 1996, 
Reinsch \etal\ 1996). Larger amplitudes are difficult to explain in this
scenario. However, the white dwarf itself varies drastically as it
expands/contracts, and in fact a flaring disk had to be assumed for 
\rxj0513\ to reduce the theoretically possible amplitude down to only 1 mag
(Hachisu \& Kato 1998).

The main features of the X-ray/optical variability of \v7\ could be  
explained by the same scenario as proposed for RX J0513.9--6951 
(Pakull \etal\ 1993, Reinsch \etal\ 1996, Southwell \etal\ 1996): 
\mdot\ variations change both the photospheric radius and 
the disk spectrum. 
If the white dwarf has a small mass than photospheric radius expansion
is reached  at 1$\times$10$^{-7}$ \msun/yr (Cassisi \etal\ 1998).
If one approximates the contraction time scale by the duration of the
mass-ejection phase (Livio 1992)
then the $\approx$50 days transition time of \v7\ into the low state 
implies a white dwarf mass of $\approx$0.8 \msun.

A major difference between the optical light curves of \rxj0513\ and 
VY Scl stars is  the amplitude between low and high states, i.e. 1 mag 
(\rxj0513) versus 3--6 mag for VY Scl stars (4 mag for \v7).
Note that the $\approx$15 eV blackbody model derived as the best fit for \v7\
corresponds to a m$_{\rm V} \approx$20 mag, i.e. several magnitudes fainter
than the observed optical low-state intensity (Fig. \ref{lc}).
Indeed, an amplitude of 4 mag can be easily accommodated by a white dwarf when
expanding from R$_{\rm WD}$ to 5 R$_{\rm WD}$. Thus, the observed large 
amplitudes in VY Scl stars could be due to a combination of both the disk 
disappearance and the white dwarf contraction.

\section{Are other VY Scl systems also SSBs?}

The finding of  luminous, supersoft X-ray emission during the optical low
state of \v7\ naturally leads to the question on how the properties of the 
VY Scl star group as a whole fit into the scenario of a SSB interpretation. 
After a short recollection of the X-ray properties of VY Scl stars
we consider the mass ratios, orbital periods, and other system properties
of VY Scl stars. These characteristics yield some clues as to
why \v7\ is a SSB and whether any other members of the VY Scl stars may
also be SSBs. The tentative interpretation we will make is that
\v7\ and possibly other VY Scl systems may represent an extension of the 
close-binary supersoft source model (van den Heuvel \etal\ 1991,
Di\,Stefano \& Nelson 1996) that appears to apply to other SSBs.

\subsection{X-ray emission of VY Scl stars at optical high state}

Not much is known about the systematics of the X-ray behaviour of VY Scl stars.
In a recent survey of available ROSAT data on all known VY Scl stars
(Greiner 1998) their X-ray emission properties  
during optical high states were found to be limited to a very narrow range of 
temperature and luminosity. Blackbody models gave the best fits to the X-ray
spectra, resulting in temperatures of 0.25--0.5 keV and luminosities in the
10$^{30}$--10$^{32}$ erg/s range. While the emission process and location is
not clear, the surprise is twofold in that VY Scl stars 
show a homogeneous X-ray spectral shape during their optical high state,
and that their X-ray spectra are distinct from other non-magnetic
cataclysmic variables (see also van Teeseling \etal\ 1996).

The upper limits derived for \v7\ during the optical high-state
(L$_{\rm X} < 10^{31}$ erg/s)
would allow for the existence of such a 0.25--0.5 keV blackbody.
If, however, such a 0.25--0.5 keV
blackbody existed in \v7\ during the optical high state 
at the same intensity level as observed in most VY Scl stars 
then it must have vanished during the transition to the optical low state 
and has been replaced by the extremely soft emission of $\approx$15 eV.

\subsection{Are all VY Scl stars supersoft X-ray binaries?}
 
\subsubsection{Group properties of VY Scl stars and how they fit the supersoft
 X-ray binary scenario}

\noindent {\bf Donor Masses and the Mass Ratio:}
In 11 of the 14 VY Scl stars, optical emission line studies indicate 
that the donor has a mass smaller than $0.5$ \msun.
In addition, the ratio between the donor's mass and that of the
white-dwarf accretor 
seems to be close to unity in some systems. Although there is
no spectral information about the donor in any SSB, and only 
indirect evidence about the mass of the white-dwarf accretor,
the mass of the donor and the mass
ratio are both typically assumed to be larger, by factors $\sim$2.   
These larger values emerge largely from theoretical models, 
and are needed to produce a mass transfer rate, $\dot m$, large enough 
($\sim 10^{-7}$ \msun/yr) to
allow the accreting matter to burn steadily on a white dwarf.

\noindent {\bf Mass of the Accreting White Dwarf:}
If the mass of the donor is small, and the mass ratio
is close to, or even smaller than unity, then the white dwarfs in
VY Scl systems may be less massive than generally considered in
models for SSBs. 
In fact, in several cases, the estimated lower limit
on the mass of the white dwarf is low enough to be consistent with
a He white dwarf. Therefore, if the lower limit is correct in any
of these cases, we may be seeing an extension (Sion \& Starrfield 1994) 
of the CO-nuclear-burning
white dwarf scenario which has formed the basic model for most SSBs so far.
The low white dwarf mass may
lead to a  larger effective radii (Vennes \etal\ 1995) and thus lower 
temperature.

\noindent{\bf White Dwarf temperatures}
White dwarf temperature estimates in VY Scl stars have been proved difficult,
and only for four systems temperatures indeed have been suggested
ranging around 50\,000 K. It has been argued by Warner (1995) that if true 
the ``high temperatures'' of these few VY Scl stars as compared to other
cataclysmic variables
are not due to ``simply radiating their original core energies''.
A detailed look at the four systems gives the following picture:
(1) DW UMa: The temperature may be incorrect because  the primary may
       be obscured by the disk (Warner 1995).
(2) V794 Aql: the temperature goes back to the fit of a Wesemael white dwarf 
    model to ``a hint of Ly-alpha absorption'', and is ``consistent with a high
    temperature'' (Szkody \etal\ 1988). 
(3) MV Lyr and TT Ari: 
    the temperatures are derived from a fit of a Wesemael white dwarf 
    model to the UV continuum, and ``is compatible with a hot (T$>$50\,000K) 
    white dwarf'' (Szkody \& Downes 1982, Shafter \etal\ 1985).
    For TT Ari, G\"ansicke \etal\ (1998) recently re-analyzed the available
    data and derive 39\,000$\pm$5\,000 K.
Thus, there is evidence -- though weak -- that the white dwarfs in 
VY Scl stars are indeed higher than in other cataclysmic variables, 
though it is presently not clear whether the temperatures are high enough
for the H burning hypothesis.

\begin{table}[th]
    \caption{Comparison of SSB and VY Scl group properties. 
    Uncertain  values are marked with a ``?''.}
    \vspace{-0.2cm}
    \begin{tabular}{ccc}
    \hline 
    \noalign{\smallskip}
     & SSBs & VY Scl stars \\
    \noalign{\smallskip} 
    \hline 
    \noalign{\smallskip}
     Mass of WD ($M_{\odot}$)    & $\sim$ 1?    & $\sim$0.5? \\
     Mass of Donor (\msun) & $\sim$ 1 -- 2? & $\sim$0.5 -- 0.7? \\
     Period (hrs)          & 6 -- 70       & 3 -- 6 \\
     kT (eV)               & 20 -- 50      & 10 -- 20 \\
     Accretion rate ($M_{\odot}$/yr) & 10$^{-7}$  & 10$^{-8}$ \\
     M$_{\rm V}$ (mag)     &  --2 -- +1    &  3 -- 5 \\
     log L (erg/s)         & 37--38      &  36 \\
     Number in Galaxy (obs) & 2          & 15 \\     
     Number in Galaxy (mod) & ~~~1000--3000~~~ & ?? \\
   \hline
   \end{tabular}
   \label{com}
\vspace{-0.1cm}
\end{table}

\noindent{\bf Orbital Periods:}
The orbital periods of VY Scl stars range from $\sim 3.2$ hours to $6$
hours, with the longest period associated with V751 Cyg. These periods
are compatible with
the orbital periods of some other SSBs, most notably 1E 0035.4--7230
with its 4.1 hr orbital period (Schmidtke \etal\ 1996). However, such 3--4 hr 
orbital periods are significantly lower than those required
in the canonical van den Heuvel \etal\ (1992) model for supersoft sources
in which a donor more massive than the white dwarf provides a mass transfer
on a thermal timescale. However, it has recently been shown that the strong
X-ray flux in supersoft sources should excite a strong wind 
($\dot M_{\rm wind} \sim 10^{-7}$ \msun/yr) from the irradiated companion
which in short-period binaries would be able to drive Roche lobe
overflow at a rate comparable to $\dot M_{\rm wind}$
(van Teeseling \& King 1998). 
The important point to note here is that even among the generally accepted,
optically identified supersoft binaries there are sources which due to their
short orbital periods are very unlikely to operate according to the
thermal-timescale mass transfer scenario of van den Heuvel \etal\ (1992).
Whether or not the above described scenario (van Teeseling \& King 1998)
to explain the  low-mass donor  supersoft systems also applies to \v7\ (or
other VY Scl stars) remains to be seen.

\noindent {\bf Accretion Rates:}
The value of $\dot m$ typically adopted in VY Scl systems (Warner 1987) is
$\dot m \sim 10^{-8} M_\odot$ yr$^{-1}$.
This is derived (using various methods) from the observed values of $M_{\rm V}$
during the optical high state. Interestingly enough, these values
may be compatible with quasi-steady burning on white dwarfs with
the low masses that seem to be indicated in some VY Scl systems. 
We note that if the disks in VY Scl stars were flared, reprocessing of the
white dwarf radiation may dominate the viscous luminosity of
the disk, thus implying accretion rates lower than the above value.

Thus, the conjecture that all VY Scl stars are SSBs is viable. 
Should it be verified, then it is likely that VY Scl stars represent an 
extension of the class of SSBs in several respects as discussed above
and summarized in Tab. \ref{com}. In addition,
independent of the value of the white dwarf mass, the evolution of 
thermal-timescale, close SSBs (those with orbital periods  $\lax 1$ day)  
predicts an epoch in which the masses have equalized, 
and the accretion rate declines (Di\thinspace Stefano \& Nelson 1996). 
The discovery of VY Scl stars
as SSBs may represent the first detections of this more quiescent phase of
the evolution of close-binary SSBs.

\subsection{Accretion disk during optical state transitions}

It is not clear what happens to the disk during the low state in VY Scl stars.
Optical spectroscopy of VY Scl stars during the very lowest states,
most notably of V794 Aql (Honeycutt \& Schlegel 1985, Szkody \etal\ 1988),
TT Ari (Shafter \etal\ 1985) and MV Lyr (Robinson \etal\ 1981), also being 
in line with our low state spectra of \v7\ (Fig. \ref{spec}), have revealed
very narrow (FWHM $\sim$ 150 km/s), strong Balmer emission and fainter HeI 
and HeII emission. In the case of TT Ari it could be shown that the phasing 
of these emission lines is shifted by 180\grad\ to that at optical 
high state, suggesting that in the very low state (1) the emission lines
are due to irradiation of the secondary, (2) there is no dominating
disk emission anymore, and (3) the presence of HeII emission lines points to 
$>$54 eV photons. Our low-state spectra reveal a FWHM $\sim$700--800 km/s 
for \v7, i.e. somewhat larger than in the above mentioned VY Scl
stars during the very lowest state.

As an alternative possibility it had been proposed that the state transition
is due to a transition from the optically thick, high \mdot\ disk
in the optical high state into an optically thin disk in the optical low 
state due to reduced mass transfer from the donor.
While this can explain the large optical amplitude between high and low state,
a simple reduction in mass transfer rate from the secondary is not
sufficient because then, after the transition to a cool state, the
disk should show outbursts like in dwarf novae (King \& Cannizzo 1998).
This has never been observed in VY Scl stars, and it has been argued
that  all disk mass must be accreted during the transition to the optical
low state. Leach \etal\ (1999) have shown that the inner disks can be kept 
in a hot state by irradiation by hot (T$\gax$ 30000K) white dwarfs.  
The latter are known to be hot in VY Scl stars,
either due to accretion heating or due to hydrogen burning as proposed here.

\subsection{Detectability of VY Scl systems in UV/EUV surveys}

We have shown that \v7\ is a very soft X-ray/EUV emitter during optical low 
states. We have therefore
considered the question whether any of the known VY Scl stars
have been detected during the EUVE survey. We have checked the second EUVE
survey catalog (Bowyer \etal\ 1996) for an entry on any of the 14 VY Scl 
stars but did not found any. We also checked the catalog of the 
cross-correlations of ROSAT all-sky survey and EUVE detections
(Lampton \etal\ 1997) which 
includes sources detected in both surveys but at lower significance
in the EUVE survey as compared to the second EUVE catalog.
These non-detections are somewhat surprising at first glance, but quite 
obvious when considering the foreground absorbing columns towards the 
VY Scl stars. For nearly all of the VY Scl stars IUE spectra have been
taken, and estimates of the extinction are available. We have used the
ROSAT count rates of the detected VY Scl stars (Greiner 1998) and their
$E(B-V)$ values ($E(B-V)$$<$0.05 implying
$N_{\rm H} < 2\times 10^{20}$ cm$^{-2}$ for V794 Aql, TT Ari, V425 Cas,
VZ Scl, VY Scl, LX Ser, DW UMa, BZ Cam, MV Lyr and $E(B-V)$=0.05 for KR Aur, 
$E(B-V)$=0.2 for V442 Oph) and estimated the predicted count rates for the
100 \AA\ band EUVE survey LEXB band. In a second step we used the
best-fit blackbody model of \v7\ as a template and applied the individual
absorbing columns of the VY Scl stars to estimate LEXB count rates. In both 
cases the derived count rates are well below the sensitivity of the EUVE survey
and the non-detection with EUVE of any VY Scl star is not in
contradiction to our suggestion of VY Scl stars possibly harboring
white dwarfs with H shell burning.

The location of the emission peak in the UV implies that 
other sources of similar kind but at possibly smaller
distance and/or less extinction should be strong EUV sources. 
Several ROSAT WFC and EUVE sources are known which so far could not be 
identified with hot white dwarfs or main-sequence stars which constitute the 
major population of EUV sources (Maoz \etal\ 1997). In particular, in analyzing
deep multicolor imaging and utilizing both color-magnitude diagrams as well as
ROSAT all-sky survey data, Maoz \etal\ (1997) conclude that 
these unidentified EUV sources are either X-ray quiet cataclysmic 
variables or a new class of objects. Given the observational result of
\v7\ being an X-ray quiet cataclysmic variable during optical high state we 
propose that nearby, hitherto unknown VY Scl stars (during optical low state)
are good candidates for these unidentified EUV sources.

An estimate of the number of unidentified EUV sources which could be VY Scl 
stars depends on several poorly known parameters such as the space density
of VY Scl stars, the duty cycle of low states, and the distance out to which 
EUV could detect emission despite interstellar absorption.
There are 4-5 known VY Scl stars within 200 pc of the Sun (Earth) while
at larger distance both the distribution perpendicular to the galactic
plane as well as absorption introduces very strong selection biases.
An even stronger selection bias is the fact that nearly all VY Scl stars
have been discovered while searching variable stars on sky patrol plates,
and thus only very few systems are known for which the bright state is
fainter than 14 mag. Given the fact that most of the known VY Scl stars 
during their optical bright state are in the 12-14 mag range, one may
expect that there is at least a similar number of VY Scl stars in the
14-16 mag range (during bright state), and thus the true number of 
VY Scl stars is higher by a factor of 2--3, implying a space density
of VY Scl stars of 3--5$\times$10$^{-7}$ pc$^{-3}$. Based on the long-term
optical light curves available for many VY Scl stars we estimate that they
spend about 5--10\% of their time in the low state.
Given the high luminosity of the UV emission it is reasonable to assume that
EUV could detect sources out to about 200 pc in the galactic plane and
even further out at high galactic latitudes. It thus seems possible that
1--3 of the unidentified EUV sources could be VY Scl stars.

\section{Conclusions}

We have observed the VY Scl star \v7\ with ROSAT during the optical 
low-state in 1997. The X-ray spectrum is very soft and the bolometric
luminosity is very high, showing that \v7\ exhibits episodes of
supersoft X-ray emission.
The anti-correlation of the X-ray and optical intensity in \v7\ resembles
the behaviour of the transient supersoft X-ray source \rxj0513.

The optical spectra taken in the optical low state are  similar to 
those of other VY Scl stars suggesting an optically thin emission region. 
However, we note the clear appearance of the high-excitation 
He\,{\sc II} 4686 \AA\ line during 
the optical low and intermediate states. This is independent evidence for an
ionization source in the system with $>$54 eV photons.

An IUE spectrum taken during an optical high-state in 1985 has a similar shape
as compared to other VY Scl stars in the same state but lacks strong
emission lines. From the 2200 \AA\ absorption dip we derive an extinction
of $E(B-V)=$0.25$\pm$0.05 and consequently estimate a distance of 
d$\approx$500 pc.
The lack of noticeable UV emission lines as well 
as the optical spectrum showing no strong orbital-phase-dependent line changes 
during the low-state spectroscopy suggest that \v7\ is viewed at a low 
inclination angle of $i$ = 30\degr$\pm$20\degr.

The absolute magnitudes in the low and high state of \v7\
are in good agreement with the expected (Warner 1987)
$M_{\rm V}$(secondary) = 16.7 -- 11.1 log P(hrs) = 8.1 mag of
a Roche-lobe filling main-sequence secondary and the luminosity of the
accretion disk $M_{\rm V}$(disk) = 5.74 - 0.259 $\times$ P(hrs) = 4.2 mag
when adopting the (uncertain) period of 6 hrs (Bell \& Walker 1980).

Our finding that  \v7\ is a transient Galactic SSB 
suggests that other VY Scl  stars may be also SSBs.
We are therefore continuing to monitor all the known VY Scl systems,
in hope of detecting luminous soft-X-ray emission from some of
the other systems during optical low states. Also other nova-like 
variables should be considered
potential candidates because of their high \mdot\ amd low white dwarf
masses similar to VY Scl stars.
Detection of X-rays from these systems, however, might turn out to be 
difficult because, if they behave like \rxj0513, would be in an expanded
state always, and thus predominantly emit in the UV band.

At this stage of our investigations,
it is important to keep an open mind about whether VY Scl
stars are SSBs, and also about the physical explanation for
SSB behavior and variability in whatever systems definitively exhibit it.
The conjecture that VY Scl stars are SSBs should now be 
subjected to further tests.     
First, if more-or-less steady nuclear burning is responsible for their
luminous supersoft X-ray emission, then during the optical high state,
the bolometric luminosity, which can be measured by a concerted 
multiwavelength campaign, using e.g. EUVE for some sources, 
should be as high as it is in the optical low state.
Second, the known VY Scl stars should be monitored closely for their
optical behaviour in order to perform further X-ray observations during
optical low states. This will allow to determine whether other VY Scl stars 
also exhibit luminous supersoft X-ray emission or whether V751 Cyg is an
exception among VY Scl stars.
In addition, there are implications such as the fact that
such low mass, low temperature white dwarfs are
difficult to detect  in external galaxies (e.g. M31), and the presence of one 
(or more, maybe 14) near us in the Galaxy may mean that previous estimates 
of the size of the population (Di\thinspace Stefano \& Rappaport 1994) 
need to be revised upward.
This would also require a re-examination of the effect of SSBs on their 
environment.

Whatever the outcome of these investigations, the
opportunity to study a larger population of local SSBs will surely
shed light on the fundamental nature of these enigmatic sources.

\begin{acknowledgements}
JG thanks Prof. J. Tr\"umper for granting ROSAT TOO time which made this
investigation only possible. We are extremely grateful to J. Mattei for 
providing very useful AAVSO data, in particular from the 1976-1979 span.
We appreciate the efforts of John Silverman and Jonathan McDowell over the 
last years in developing the code to compute the HRI detector response 
matrices. We thank the referee for a thorough reading and
helpful comments which substantially improved this paper.
JG is supported by the German Bundesmi\-ni\-sterium f\"ur Bildung, 
Wissenschaft, Forschung und Technologie (BMBF/DLR) under contract No. 
FKZ 50 QQ 9602 3, and GT acknowledges funding by CONACYT grant 25454-A. 
The \ros\, project is supported by BMBF/DLR and the
Max-Planck-Society. This research has made use of the Simbad database, 
operated at CDS, Strasbourg, France. Fig. \ref{fc} is based on photographic 
data of the National Geographic Society -- Palomar
Observatory Sky Survey (NGS-POSS) obtained using the Oschin Telescope on
Palomar Mountain.  The NGS-POSS was funded by a grant from the National
Geographic Society to the California Institute of Technology.  The
plates were processed into the present compressed digital form with
their permission.  The Digitized Sky Survey was produced at the Space
Telescope Science Institute under US Government grant NAG W-2166.
\end{acknowledgements}


\begin{thebibliography}{}

\bibitem[]{aaaa97} Alcock C., Allsman R.A., Alves D., \etal\, 1997,
   MNRAS 286, 483

\bibitem[]{a73} Allen C.W., 1973, Astrophysical Quantities, Athlone Press, 
   London

\bibitem[]{bw80} Bell M., Walker M.F., 1980, BAAS 12, 63

\bibitem[]{b97} Bergh\"ofer Th., 1997 (priv. comm.)

\bibitem[]{bs81} Bopp B.W., Stencel R.E., 1981, ApJ 247, L131

\bibitem[]{bllwjm96} Bowyer S., Lampton M., Lewis J., Wu X., Jelinsky P.,
   Malina R.F., 1996, ApJ Suppl. 102, 129

\bibitem[]{bm73} Burrell J.F., Mould J.R., 1973, PASP 85, 627

\bibitem[]{cit98} Cassisi S., Iben I.Jr., Tornambe A., 1998, ApJ 496, 376

\bibitem[]{cjn81} Cordova F.A., Jensen K.A., Nugent J.J., 1981, MNRAS 196, 1

\bibitem[]{dhkz97} David L.P., Harnden F.R., Kearns K.E., Zombeck M.V., 1997,
The ROSAT HRI Calibration Report, available from the URL addresses
http:/$\!$/hea\-www.harvard.edu/\-rosat/rsdc\_www/HRI\_CAL\_REPORT/hri.html

\bibitem[]{dsr94} DiStefano R., Rappaport S., 1994, ApJ 437, 733

\bibitem[24]{dn96} Di\thinspace Stefano R., Nelson L.A., 1996, in Supersoft 
X-ray Sources,   ed. J. Greiner, Lecture Notes in Phys. 472, Springer, p. 3

\bibitem[]{dbyf94} Dobashi K., Bernard J.-P., Yonekura Y., Fukui Y., 1994,
   ApJS 95, 419

\bibitem[]{dhpw95} Downes R., Hoard D.W., Szkody P., Wachter S., 1995, AJ 110,
   1824

\bibitem[]{du80} Drake S.A., Ulrich R.K., 1980, ApJ Suppl. 42, 351

\bibitem[]{ecm93} Echevarria J., Costero R., Michel R., 1993, A\&A 275, 201

\bibitem[]{fuji82} Fujimoto M.Y., 1982, ApJ 257, 767

\bibitem[]{gaen98} G\"ansicke B., Sion E.M., Beuermann K. \etal\ 1998, 
  A\&A (to be subm.)

\bibitem[]{grei96} Greiner J. (Ed.), 1996, Supersoft X-ray Sources,
  Lecture Notes in Phys. 472, Springer

\bibitem[]{gsho96} Greiner J., Schwarz R., Hasinger G., Orio M., 1996, 
      A\&A 312, 88

\bibitem[]{g98}  Greiner J., 1998, A\&A 336, 626

\bibitem[]{gds98} Greiner J., Di\,Stefano R., 1998, in ``Highlights in
X-ray Astronomy'', eds. B. Aschenbach et al., Garching, June 1998, 
MPE Report (in press) (astro-ph/9810019)

\bibitem[]{hk98} Hachisu I, Kato M., 1998, ApJ (in press)

\bibitem[]{h58} Herbig G.H., 1958, ApJ 128, 259

\bibitem[]{hr72} Herbig G.H., Rao N.K., 1972, ApJ 174, 401

\bibitem[]{hs97} Hoard D.W., Szkody P., 1997, ApJ 481, 433

\bibitem[]{hs85} Honeycutt R.K., Schlegel E.M., 1985, PASP 97, 1189

\bibitem[]{jhtvbv90} Jetsu L., Huovelin J., Tuominen I., Vilhu O., Bopp B.W.,
  Piirola V., 1990, AA 236, 423

\bibitem[]{king97} King A.R., 1997, MNRAS 288, L16

\bibitem[]{kc98} King A.R., Cannizzo J.K., 1998, ApJ 499, 348

\bibitem[]{kl98} Knigge C., Livio M., 1998, MNRAS 297, 1079

\bibitem[]{kra64} Kraft R.P., 1964, in ``First Conf. on Faint Blue Stars'', 
  ed. W.J. Luyten (Minneapolis: Univ. of Minnesota), p. 100

\bibitem[]{ldous91} la Dous C., 1991, AA 252, 100

\bibitem[]{ldous93} la Dous C., 1993, in ``Cataclysmic variables and Related 
  Objects'', Eds. Hack and La Dous, NASA SP-507, p. 107


\bibitem[]{llsbvlw97} Lampton M., Lieu R., Schmitt J.H.M.M., Bowyer S.,
   Voges W., Lewis J., Wu X., 1997, ApJ Suppl. 108, 545

\bibitem[]{lando92} Landolt A.U., 1992, AJ 104, 340

\bibitem[]{lhk99} Leach R., Hessman F.V., King A.R., Stehle R., Mattei J., 
  1999, MNRAS (in press)

\bibitem[]{livio92} Livio M., 1992, ApJ 393, 516

\bibitem[]{lp94} Livio M., Pringle J.E., 1994, ApJ 427, 956

\bibitem[]{mos97} Maoz D., Ofek E.O., Shemi A., 1997, MNRAS 287, 293

\bibitem[]{m58} Martynov D.Y., 1958, Perem. Zvezdy 11, 170

\bibitem[]{mk58} Martynov D.Y., Kholopov P.N., 1958, Perem. Zvezdy 11, 222

\bibitem[]{mzb97} Munari U., Zwitter T., Bragaglia A., 1997, A\&AS 122, 495

\bibitem{nk80} Neckel Th., Klare G., 1980, A\&A Suppl. 42, 251

\bibitem[]{osa96} Osaki Y., 1996, PASP 108, 39

\bibitem[]{pmb93} Pakull M.W., Motch C., Bianchi L., \etal\
              1993, A\&A 278, L39

\bibitem[]{patt84} Patterson J., 1984, ApJS 54, 443

\bibitem[]{ps95} Predehl P., Schmitt J.H.M.M., 1995, A\&A 293, 889

\bibitem[]{pres98} Prestwich A.H., Silverman A., McDowell J., Callanan P., 
 Snowden S., 1998 (in prep.); see also the URL
  address  http:/$\!$/heasarc.gsfc.nasa.gov/docs/rosat/hrispec.html

\bibitem[]{rtba96} Reinsch K., van Teeseling A., Beuermann K., Abbott T.M.C., 
     1996, A\&A 309, L11

\bibitem[]{rnk74} Robinson E.L., Nather R.E., Kiplinger A., 1974, PASP 86, 401

\bibitem[]{rbcc81} Robinson E.L., Barker E.S., Cochran A.L., Cochran W.D., 
    Nather R.E., 1981, ApJ 251, 611

\bibitem{rswp98} Rodr\'{\i}guez-Pascual P., Schartel N., Wamsteker W., 
    P\'erez A. 1998, in ``UV Astrophysics beyond the IUE Final Archive'', 
   Eds. Gonz\'alez-Riestra \etal, ESA SP-413, p. 731

\bibitem{rbhsga79} Rothschild R., Boldt E., Holt S., \etal\
  1979,  Space Sci. Instr. 4, 269

\bibitem{sht93} Schaeidt S., Hasinger G., Tr\"umper J., 1993, A\&A 270, L9

\bibitem{scmhc96} Schmidtke P.C., Cowley A.P., McGrath T.K., Hutchings J.B.,
   Crampton D., 1996, AJ 111, 788

\bibitem{schr94} Schrijver C.J., 1994, in Cool Stars, Stellar Systems and the
     Sun, ed. J.-P., Caillault, ASP Conf. Ser. 64, ASP, San Francisco, p. 328

\bibitem{sslpd85} Shafter A.W., Szkody P., Liebert J., Penning W.R.,
    Bond H.E., Grauer A.D., 1985, ApJ 290, 707

\bibitem[]{ss94} Sion E.M., Starrfield S.G., 1994, ApJ 421, 261

\bibitem{slc96} Southwell K.A., Livio M., Charles P.A., O'Donoghue D., 
   Sutherland W.J., 1996, ApJ 470, 1065

\bibitem{szk81} Szkody P., 1981, PASP 93, 456

\bibitem{sd82} Szkody P., Downes R.A., 1982, PASP 94, 328

\bibitem{sdm88} Szkody P., Downes R.A., Mateo M., 1988, PASP 100, 362

\bibitem{trwsn91} Thorstensen J.R., Ringwald F.A., Wade R.A., Schmidt G.D.,
   Norsworthy J.E., 1991, AJ 102, 272

\bibitem{vdhbnr92} van den Heuvel E.P.J., Bhattacharya D., Nomoto K., 
    Rappaport S.A.,  1992, A\&A 262, 97

\bibitem{tbv96} van Teeseling A., Beuermann K., Verbunt F., 1996, A\&A 315, 467

\bibitem{trh97} van Teeseling A., Reinsch K., Hessman F.V., Beuermann K.,
   1997,  A\&A, L41

\bibitem{tk98} van Teeseling A., King A.R., 1998, A\&A 1998, A\&A 338, 957

\bibitem[23]{vfb95} Vennes S., Fontaine G.,  Brassard P., 1995,
     A\&A 296, 117

\bibitem[]{w87} Warner B., 1987, MNRAS 227, 23

\bibitem[]{w95} Warner B., 1995, Cataclysmic Variable Stars, Cambridge 
   Astrophys. Ser. 28, Cambridge Univ. Press

\bibitem[]{ww63} Wenzel W., 1963, Mitt. Ver. Sterne Nos. 754 and 755

\bibitem[]{wtwvp96} Wheatley P.J., van Teeseling A., Watson M.G., Verbunt F.,
  Pfeffermann E., 1996, MNRAS 283, 101

\end{thebibliography}
\end{document}